\documentclass{emulateapj}
\usepackage{graphicx}
\usepackage{color}

\shortauthors{Carter et al.\ 2010}
\shorttitle{Sixteen Transits of GJ~1214b}

\begin{document}

% ------------------------------------------------------------------------
% New commands
%
\def\ltsima{$\; \buildrel < \over \sim \;$}
\def\lsim{\lower.5ex\hbox{\ltsima}}
\def\gtsima{$\; \buildrel > \over \sim \;$}
\def\gsim{\lower.5ex\hbox{\gtsima}}
                                                                                          
% -------------------------------------------------------------------------
%

\bibliographystyle{apj}

\title{The Transit Light Curve Project.~\ron{XIII}.~Sixteen
Transits of the Super-Earth GJ~1214\lowercase{b}\footnote{T\lowercase{his paper includes data gathered with the 6.5 meter}~M\lowercase{agellan telescopes located at}~L\lowercase{as}~C\lowercase{ampanas}~O\lowercase{bservatory,}~C\lowercase{hile.}}}

\author{
Joshua A.\ Carter\altaffilmark{1,2,4},
Joshua N.\ Winn\altaffilmark{1},
Matthew J.\ Holman\altaffilmark{2},
Daniel Fabrycky\altaffilmark{2,3,4},\\
Zachory K.\ Berta\altaffilmark{2},
Christopher J.\ Burke\altaffilmark{2},
Philip Nutzman\altaffilmark{2,3}
}

\newcommand{\ron}{}
\newcommand{\new}{}
%\newcommand{\remove}{\bf \sout}

% \journalinfo{Draft version, October 22, 2010}
% \slugcomment{To be submitted to ApJ}

\altaffiltext{1}{Department of Physics, and Kavli Institute for
  Astrophysics and Space Research, Massachusetts Institute of
  Technology, Cambridge, MA 02139}

\altaffiltext{2}{Harvard-Smithsonian Center for Astrophysics, 60
 Garden St., Cambridge, MA 02138}

\altaffiltext{3}{Department of Astronomy and Astrophysics, University of California, Santa Cruz, CA 95064}

\altaffiltext{4}{Hubble fellow}

\begin{abstract}

We present optical photometry of 16 transits of the super-Earth GJ~1214b, allowing us to refine the system parameters and search for additional planets via transit timing. Starspot-crossing events are detected in two light curves, and the star is found to be variable by a few percent. Hence, in our analysis, special attention is given to systematic errors that result from star spots. The planet-to-star radius ratio is $0.11610\pm 0.00048$, subject to a possible upward bias by a few percent due to the unknown spot coverage. Even assuming this bias to be negligible, the mean density of planet can be either $3.03\pm 0.50$~g~cm$^{-3}$ or $1.89\pm 0.33$~g~cm$^{-3}$, depending on whether the stellar radius is estimated from evolutionary models, or from an {\new{empirical mass-luminosity relation combined with}} the light curve parameters. One possible resolution is that the orbit is eccentric ($e \approx 0.14$), which would favor the higher density, and hence a much thinner atmosphere for the planet. The transit times were found to be periodic within about 15~s, ruling out the existence of any other super-Earths with periods within a factor-of-two of the known planet.

\end{abstract}
		

\section{Introduction}

The recently discovered planet GJ~1214b (Charbonneau et al.~2009) is the smallest known exoplanet for which the mass, radius, and atmospheric properties are all possible to study with current technology. It is therefore a keystone object in the theory of planetary interiors and atmospheres, and has been welcomed as the harbinger of the ``era of super-Earths'' (Rogers \& Seager~2010).

The planet's discoverers estimated the mass and radius of GJ~1214b to be $6.55\pm 0.98$~$M_\oplus$ and $2.68\pm 0.13$~$R_\oplus$, giving a mean density of $1.87\pm 0.40$~g~cm$^{-3}$ (Charbonneau et al.~2009). This is such a low density that it would seem impossible for the planet to be solid with only a thin, terrestrial-like atmosphere. Rather, it seems necessary to have a thick gaseous atmosphere, probably composed of hydrogen and helium but possibly also of carbon dioxide or water (Charbonneau et al.~2009, Rogers \& Seager 2010, Miller-Ricci \& Fortney 2010).

In this paper, we report on observations of additional transits of GJ~1214b. As in other papers in the Transit Light Curve (TLC) series (Holman et al.~2006, Winn et al.~2007), one of our goals was to refine the basic system parameters, and thereby allow for more powerful discrimination among models of the planet's interior and atmosphere.  Another goal was to check for any non-periodicity in the transit times, as a means of discovering other planets in the system, through the method of Holman \& Murray (2005) and Agol et al.~(2005). Super-earths have frequently been found in pairs or even triples in compact arrangements (Lo Curto et al.~2010), and it would be interesting to know if GJ~1214b is another such example.

This paper is organized as follows. Section \ref{sec:obsred} describes the observations and data reduction.  Section~\ref{sec:model} presents the light curve model, taking into account the effects of starspots. Section~\ref{sec:analysis} discusses the method by which we estimated the model parameters and their confidence intervals. Section~\ref{sec:radrat} discusses the results for the planet-to-star radius ratio. Section~\ref{sec:radplanet} presents two different methods for determining the stellar radius (and hence the planetary radius), which give discrepant results. Some possible resolutions of this discrepancy are discussed.  Section~\ref{sec:timing} presents our analysis of the measured transit times, and constraints on the properties of a hypothetical second planet.  Finally, in Section~\ref{sec:disc}, we discuss the implications of our analysis on the understanding of GJ~1214b and more broadly on M dwarf transit hosts.

\section{Observations and Data Reduction} \label{sec:obsred}

Our data were gathered during the 2009 and 2010 observing seasons. Thirteen transits were observed with the 1.2m telescope at the Fred L.\ Whipple Observatory (FLWO) on Mount Hopkins, Arizona, using Keplercam and a Sloan $z'$ filter. {\new{The blue end of the bandpass
was set by the filter (approx.\ 0.85~$\mu$m) and the red end
by the quantum efficiency of the CCD (ranging from nearly 100\% at 0.75~$\mu$m to
10\% at 1.05~$\mu$m).}}
The first two of the FLWO transits were already presented by Charbonneau et al.\ (2009); those data have been reanalyzed for this work. Another three transits were observed with the 6.5m Magellan ({\new{Baade}}) telescope at the Las Campanas Observatory in Chile, with the MagIC and IMACS cameras and a Sloan $r'$ filter {\new{(approx.\ 0.55-0.70~$\mu$m)}} 

We used standard procedures for bias subtraction and flat-field division. Circular aperture photometry was performed for GJ~1214 and several comparison stars (5-6 for the FLWO images, and 2 for the Magellan images). Circular annuli centered on each star were used to estimate the background level. The flux of GJ~1214 was divided by the sum of the fluxes of comparison stars, to produce a relative flux time series for GJ~1214. The radii of the apertures and annuli were varied in order to seek those values that minimize the scatter in the residuals, while also producing a minimal level of {\new{time-correlated}} noise, as determined by the wavelet-based algorithm described by Carter \& Winn (2009). Because the amplitude of the correlated component was always $\lesssim 1\%$ of that of the white component, in what follows we assume the noise to be uncorrelated in time. Each measurement was associated with a time stamp given by the midexposure time expressed in the BJD$_{\rm TDB}$ system (Eastman et al.~2010).

\begin{figure*}[th] %  figure placement: here, top, bottom, or page
   \centering
     %\epsscale{1.15}
      \plotone{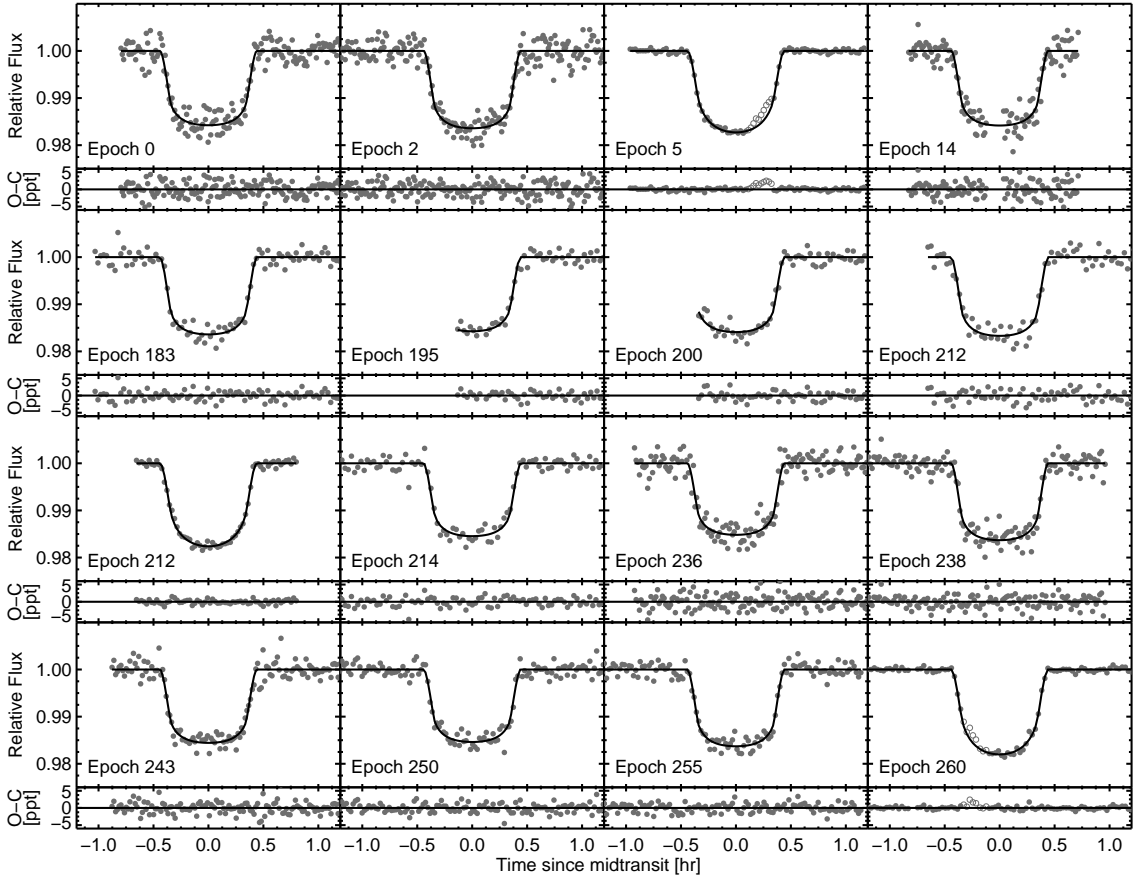}
      \caption{GJ~1214 transit light curve photometry.  Each panel shows transit light data (after correcting for airmass variation), the best-fitting model, and the residuals (observed~$-$~calculated, in parts per thousand).  Unfilled circles represent the
suspected ``spot anomalies'' that were assigned zero weight in the fitting process.  The data from epochs 5, 212, and 260 are based
on $r'$-band observations with the Magellan Clay 6.5m telescope. The other data are based on $z'$-band observations
with the FLWO 1.2m telescope.}
   \label{fig:bigplots}
\end{figure*}

\begin{figure*}[th] %  figure placement: here, top, bottom, or page
   \centering
     %\epsscale{1.15}
      \plotone{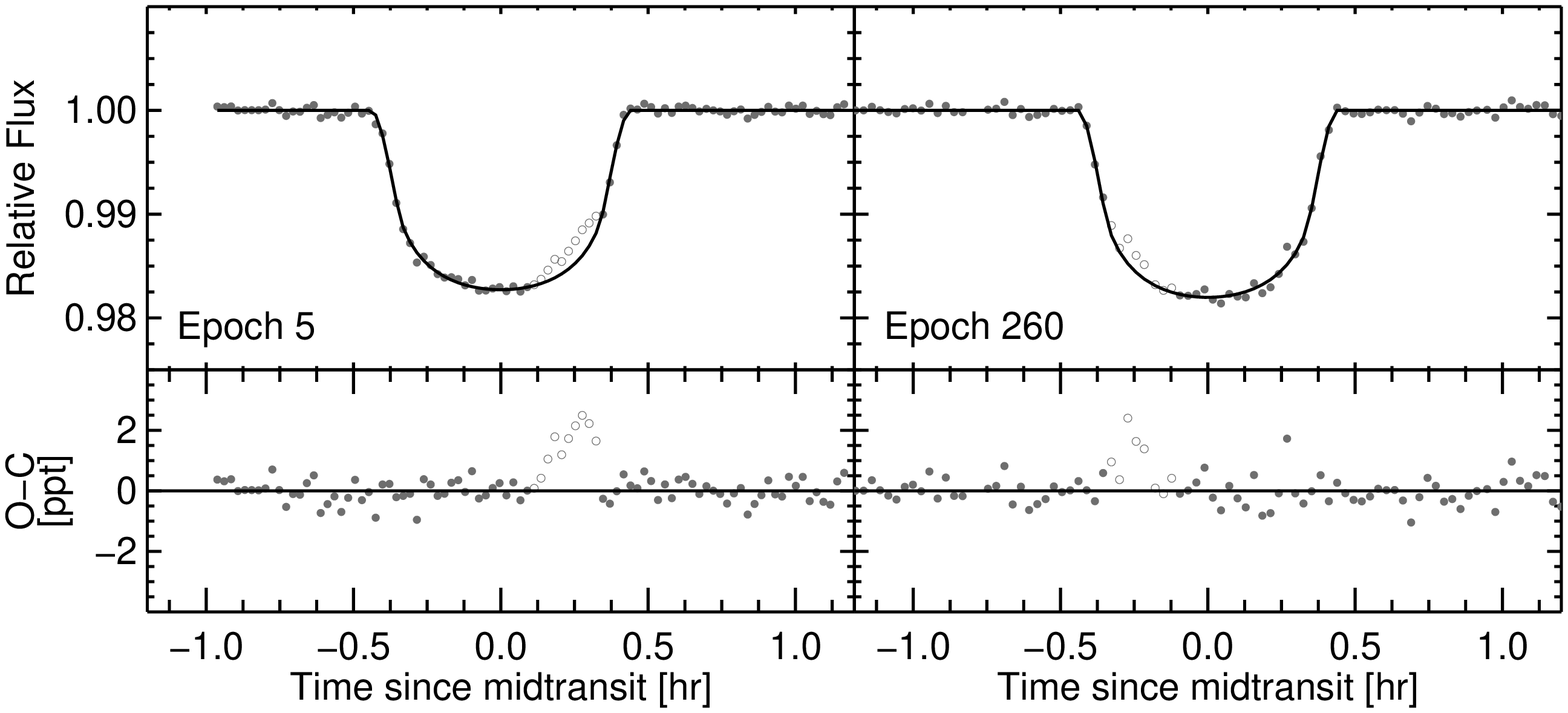}
      \caption{Close-ups of the two light curves with suspected spot anomalies. Both light curves are based on
$r'$-band observations with the Magellan Clay 6.5m telescope. The plotting conventions are the same
as in Figure~\ref{fig:bigplots}.}
   \label{fig:closeup}
\end{figure*}

Fig.~\ref{fig:bigplots} shows the light curves, after correcting for differential airmass extinction as described in \S~\ref{sec:analysis}.  Fig.~\ref{fig:closeup} gives a clearer view of two of the Magellan light curves (epochs 5 and 260), which show evidence for anomalous brightening events during the transit. The data points corresponding to these events are shown with unfilled circles, rather than gray circles. Correlations were sought between the flux observed during these events, and various image statistics such as the median X and Y pixel position, the shape parameters of the point spread function, and the background level, but none were found. We interpret each of those events as the passage of the planet in front of a dark starspot, as has been observed by Rabus et al.~(2009) and others for different transiting systems.

\section{Transit light curve model} \label{sec:model}

The evidence for starspots on GJ~1214 caused us to make a few changes to the usual models for transit photometry. Cool starspots on the visible hemisphere of the star can produce two different effects: when they are on the transit chord, they cause brightening events in the light curve such as those we observed; and when they are away from the transit chord, they reduce the overall brightness of the star and increase the fractional loss of light due to the planet. The latter effect causes the planet to appear larger than it is in reality, by producing a deeper transit.

\subsection{Loss of light due to starspots} \label{sec:model:spots}

During a transit, the flux received from a spotted star can be written
\begin{equation}
F(t) = F_0[1-\epsilon(t)] - \Delta F(t),
\label{eq:spot-model}
\end{equation}
where $F_0$ is the flux that would be received from an unspotted and untransited star, $\epsilon(t)$ is the fractional loss of light due to starspots, and $\Delta F(t)$ is the flux that is blocked by the planet. The factor $\epsilon(t)$ changes on the timescale of the stellar rotation period ($\sim$~months), while $\Delta F(t)$ changes on the much shorter timescale of the transit ($\sim$~hours).

It is convenient to normalize the light curve by the flux measured
just outside of the transit,
\begin{equation}
f(t) \equiv 1 - \frac{1}{1-\epsilon} \frac{\Delta F(t)}{F_0},
\end{equation}
where $\epsilon$ is taken here to be a constant for the duration of the transit.  The quantity $\Delta F(t)/F_0$ is the fractional loss of light due to the transit, which can be calculated in the usual way. In this work we use the formulas of Mandel \& Agol (2002) for a quadratic limb-darkening law.  Thus, our model adds one new parameter $\epsilon$ specific to each transit. A similar model was used for Corot-2b by Czesla et al.~(2009).

\begin{figure*}[t] %  figure placement: here, top, bottom, or page
\centering
\epsscale{1.15}
\plottwo{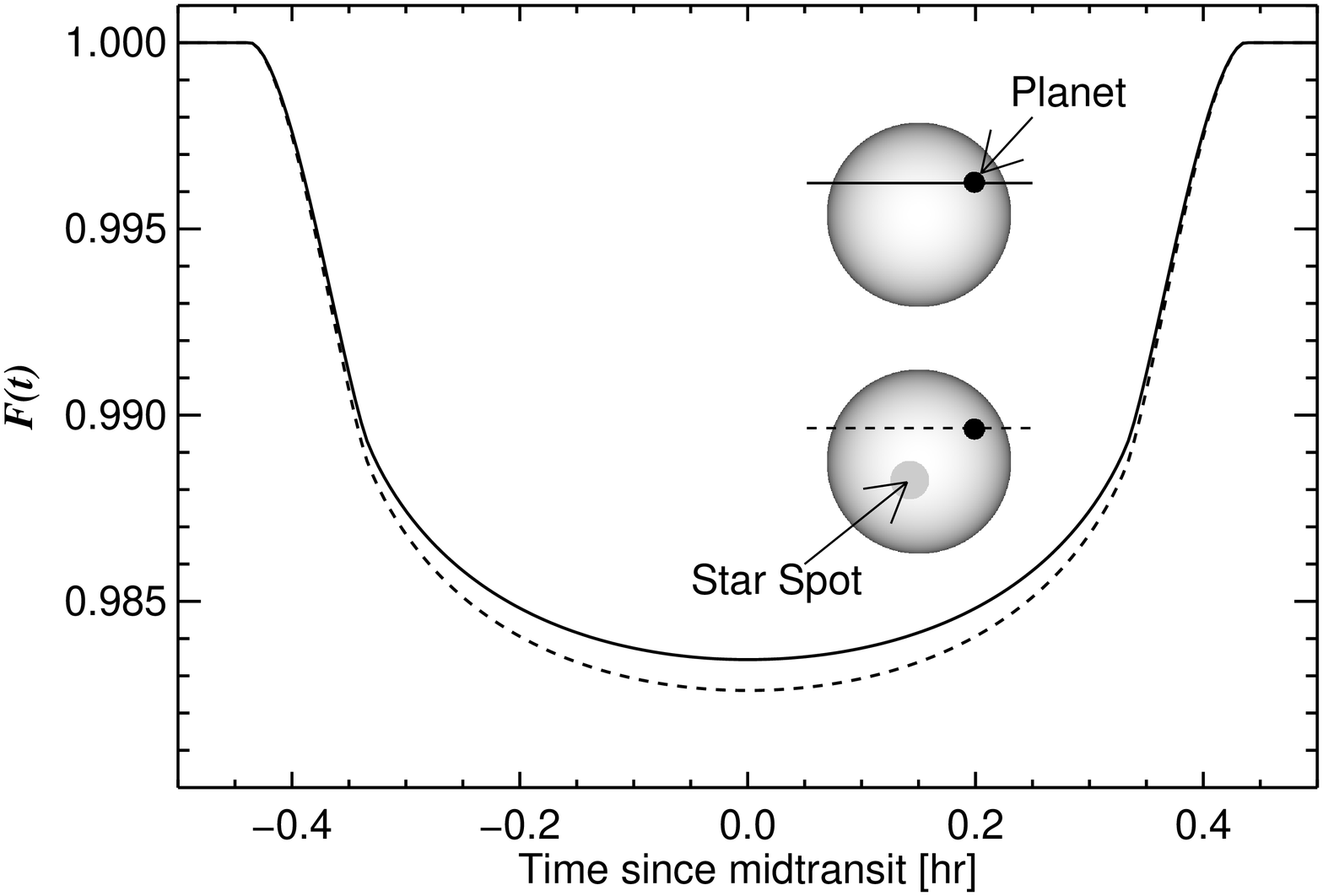}{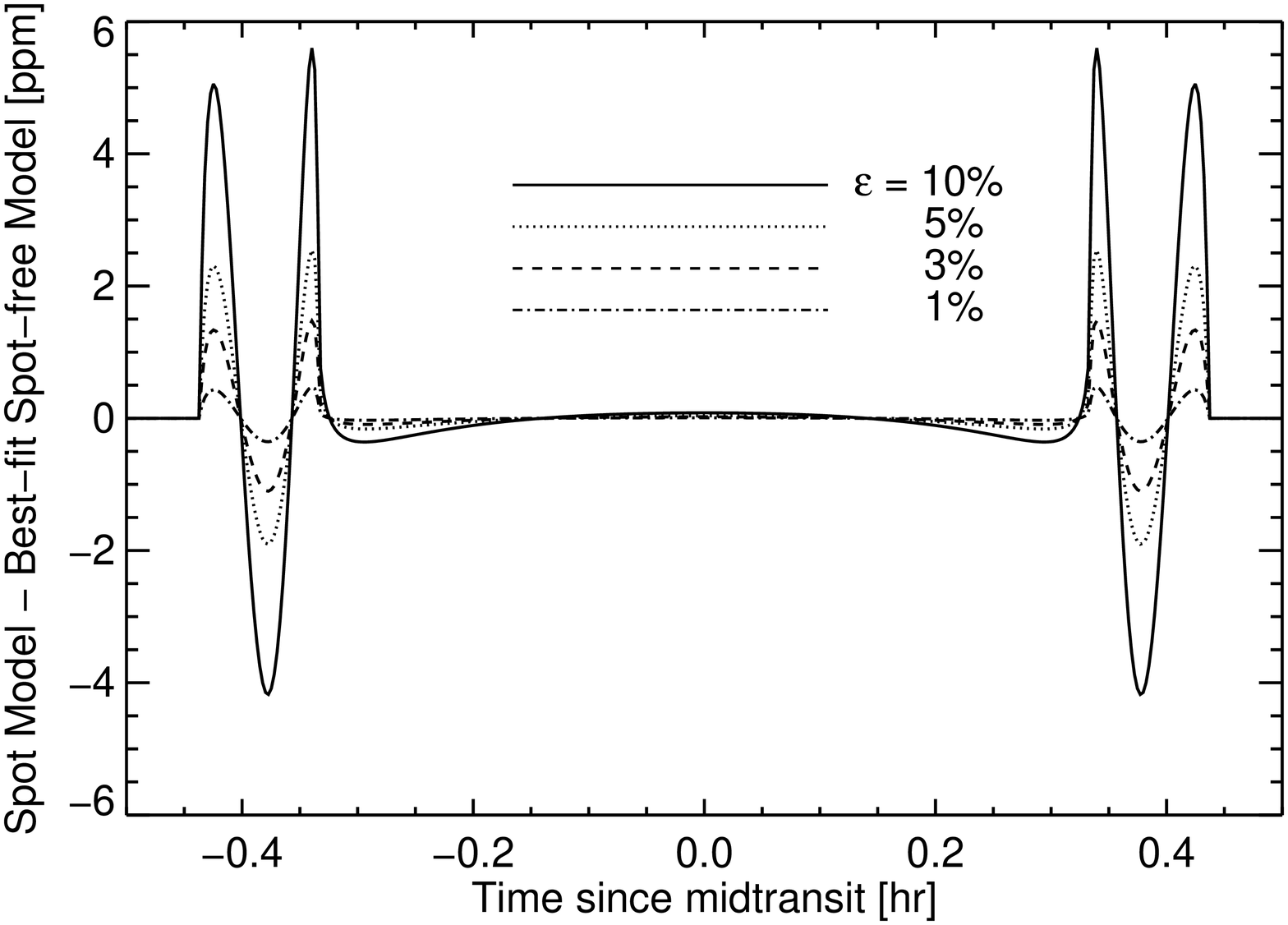}
\caption{The effect of {\it untransited} starspots.  {\it Left.}---The light curves of a planet transiting a star with (dashed line) and without (solid line) a starspot on the visible disk [$\epsilon = 5\%$ {\new{in Eqn.~(\ref{eq:spot-model})}}].  In both cases the out-of-transit flux was set equal to unity.  {\it Right.}---Residuals between the light curves and the best-fitting spot-free light curve. }
\label{fig:spot}
\end{figure*}

With infinite photometric precision, $\epsilon$ could be determined from a single transit light curve based on the slight differences in the ingress or egress data as compared to a spot-free model.  However, the differences are typically no greater than a few parts per million, and will be difficult to detect in practice (see the right panel of Fig.~\ref{fig:spot}). Consequently, there is a strong degeneracy between $\epsilon$ and the planet-to-star radius ratio $R_p/R_\star$, in the sense that increasing either parameter results in a deeper transit with no other observable changes. If the spots produce an overall darkening ($\epsilon \geq 0$) then the measured transit depth $\delta$ (corrected to remove the effects of limb darkening) can only be used to place \ron{an} upper bound on the radius ratio:
\begin{eqnarray}
	\frac{R_p}{R_\star} = \sqrt{ \delta (1-\epsilon)} & ~\leq \sqrt{\delta}.
\end{eqnarray}
If the spot coverage is time-variable, then the transit depth will vary from event to event. One may hope that the shallowest observed transit corresponds to a time when the visible disk was nearly free of spots, thereby allowing the planet-to-star radius ratio to be measured with minimal bias. However, there is no guarantee that the spot coverage ever falls to zero.

\subsection{Spot crossings during transit} \label{sec:spotcross}

When the planet transits a cool spot, the fractional loss of light temporarily decreases, and a brightening is observed in the light curve. The duration and amplitude of the brightening event depends on the size of the spot and the intensity contrast with the unspotted stellar photosphere. If the spot is approximately the same size as the planet, and is completely occulted, then the brightening event will have a duration of approximately twice the transit ingress duration. For a circular spot, the amplitude of the brightening event is
\begin{eqnarray}
\delta_{\rm spot} & = &
\left(\frac{R_s}{R_\star}\right)^2\left(1-\frac{{\cal I}_s}{{\cal I}_\star}\right)
\end{eqnarray}
where $R_s$ and ${\cal I}_s$ are the radius and mean intensity of the spot, and ${\cal I}_\star$ is the mean intensity that the photosphere would have at that location in the absence of spots.

Rather than modeling the spots {\new{(e.g., Wolter et al. 2009)}} and fitting for the brightening events, we chose the simple approach of identifying the brightening events visually, and assigning those data zero weight in the fitting process. For the Magellan data, the identification could be performed without much ambiguity, and the ``excised'' data are shown in Figs.~1 and 2.  For the FLWO data, no brightening events were obvious to the eye. This is probably due to a combination of the lower precision of the FLWO data, and the weaker contrast between the spots and the photosphere at the longer wavelengths of the FLWO observations ($z'$ band as compared to $r'$ band).

In particular, the anomalies we observed at epochs 5 and 260 have a duration of approximately twice the ingress duration, and therefore the typical spots are likely to be comparable in size to the planet. The observed amplitudes of 2--3 ppt (parts per thousand) give a constraint on the spot size and the intensity contrast. The lowest possible intensity contrast is obtained for the largest possible spots, with $R_s \approx R_p$. In that case, assuming the unspotted photosphere and the spots to be described by blackbody spectra, the spot temperature would be $T_s \approx 2900$~K as compared to the photospheric temperature of 3026~K (Charbonneau et al.\ 2009). Had we observed those same events in the $z'$ band, the brightening amplitude would have been $\approx$1.5~ppt, just below the 1$\sigma$ level of the noise in each data point. Smaller spots would need to have a greater intensity contrast, leading to a greater temperature difference and an even smaller predicted amplitude in the $z'$ band.

In this light it is not surprising that we did not detect similar starspot events in our $z'$-band data. In what follows we assume that the $z'$-band data are unaffected by spot-crossing events. Of course it is possible that smaller or less luminous starspots were transited during our observations, and that the transit depths are ``filled in'' to some degree by brightening events that cannot be identified visually in the light curves. However, the reasoning in the preceding paragraph suggests that this effect is minor.  Furthermore, we did not find any evidence for time correlated noise in our model residuals.

\subsection{Using transit durations to estimate the radius ratio}

As discussed, the effect of dark spots on the visible disk of the star is to increase the fractional loss of light during the transit, causing the planet to appear larger than it is in reality. This may be counteracted to some degree by the ``filling in'' of the transit due to spot occultations, if those events cannot be recognized and excised from the data. This caused us to wonder what one could still learn about the planet if the transit depth cannot be trusted.

The transit light curve provides three primary observables: the total transit duration $T_{\rm tot}$, the ingress or egress duration $\tau$, and the depth $\delta$. If one also knows the orbital period $P$, eccentricity $e$, and argument of pericenter $\omega$, then one can translate the transit observables into three parameters more directly related to the star-planet system.  One possible choice of three ``physical'' parameters is the planet-to-star radius ratio $R_p/R_\star$, the orbital inclination $i$, and the mean stellar density $\rho_\star$ (see, e.g., Carter et al.~2008 or Winn~2010).  Usually, the value of $\rho_\star$ derived in this way is used to improve the characterization of the host star (see, e.g., Sozzetti et al.~2007, Holman et al.~2007).  However, if one is willing to adopt a value for $\rho_\star$ based on external information (such as the star's observed luminosity and spectrum), then the two observables $T_{\rm tot}$ and $\tau$ are sufficient to specify the other two physical parameters $i$ and $R_p/R_\star$.

Hence one may derive $R_p/R_\star$ based on $T_{\rm tot}$ and $\tau$ alone, without any reference to $\delta$. The result of straightforward algebra is
\begin{eqnarray}
\frac{\left(T_{\rm total}-\tau\right)\tau}{4 \left[3 P/(8 \pi^2 G)\right]^{2/3}} & = & \left(\frac{R_p}{R_\star} \right) \rho_\star^{-\frac{2}{3}} \frac{\left(1+e \sin \omega\right)^2 }{1-e^2} \nonumber \\
&& + O\left[\frac{\rho_p}{\rho_\star} \left(\frac{R_p}{R_\star}\right)^4\right],
\label{eq:duration-radratio}
\end{eqnarray}
where $G$ is the gravitational constant. For cases where star spots are likely to be present, this ``duration-based'' method of deriving $R_p/R_\star$ is a useful alternative to the usual ``depth-based'' method, and it may be more accurate in situations when $\rho_\star$ is well constrained. The duration-based method is immune to the bias that is caused by starspots outside of the transit chord, although one must still be wary of starspot anomalies that occur during ingress or egress. As we will describe, for GJ~1214 we considered both depth-based and duration-based methods to estimate $R_p/R_\star$.

\section{Transit light curve analysis} \label{sec:analysis}

We modeled the data that are plotted in Fig.~\ref{fig:bigplots} using the approach represented by Equation~(\ref{eq:spot-model}). Each transit was assigned a parameter $\epsilon$, representing the fractional loss of light due to spots on that night. Since a degeneracy prevents the determination of all of the $\epsilon$ values along with $R_p/R_\star$, we fixed $\epsilon=0$ for the transit with the shallowest depth (epoch 236, UT~2010~June~5) and allowed all the others to be free parameters. This is equivalent to assuming that the visible stellar disk had no spots on that particular night. The results can be scaled appropriately for any assumed value of the spot coverage on that night.

We used the Mandel \& Agol (2002) equations to calculate $\Delta F/F_0$, assuming a quadratic limb-darkening law. The orbit was assumed to be circular. The time of conjunction $t_c$ for each transit was allowed to be a free parameter, but for the limited purpose of computing the scaled orbital distance $a/R_\star$, we assumed the orbital period to be $1.5803925$ days (Charbonneau et al.~2009). We also allowed the out-of-transit magnitude on each night to be a linear function of airmass, in order to allow for color-dependent differential extinction between GJ~1214 and the bluer comparison stars.

All together, there were 70 model parameters: $R_p/R_\star$, $a/R_\star$, $i$, $u_r$ and $v_r$ (the $r'$-band limb darkening coefficients), $u_z$ and $v_z$ (the $z'$-band limb darkening coefficients), $\epsilon_j$ (for $j=1$ to 15), $t_{c,j}$ (for $j=1$ to 16), and two constants specifying the linear airmass correction for each of {\new{the}} 16 transits. We determined the posterior probability distributions for these parameters with a Markov Chain Monte Carlo (MCMC) algorithm, using Gibbs sampling and Metropolis-Hastings stepping {\new{(see, e.g.,
Tegmark et al.\ 2004, Holman et al.\ 2006)}}.  The likelihood was taken to be $\exp(-\chi^2/2)$, with \begin{eqnarray}
	\chi^2 & = & \sum_k \frac{\left[F_k - {\cal F}(t_k)\right]^2}{\sigma^2_k} \label{eq:chi2}
\end{eqnarray}
where $F$ is the measured flux, ${\cal F}$ is the calculated flux, and $\sigma_k$ is root-mean-square of the residuals from the best-fitting model specific to each night. The data from the suspected spot-induced anomalies during epochs 5 and 260 were excluded from this sum.

\begin{figure*}[t] %  figure placement: here, top, bottom, or page
   \centering
           \plotone{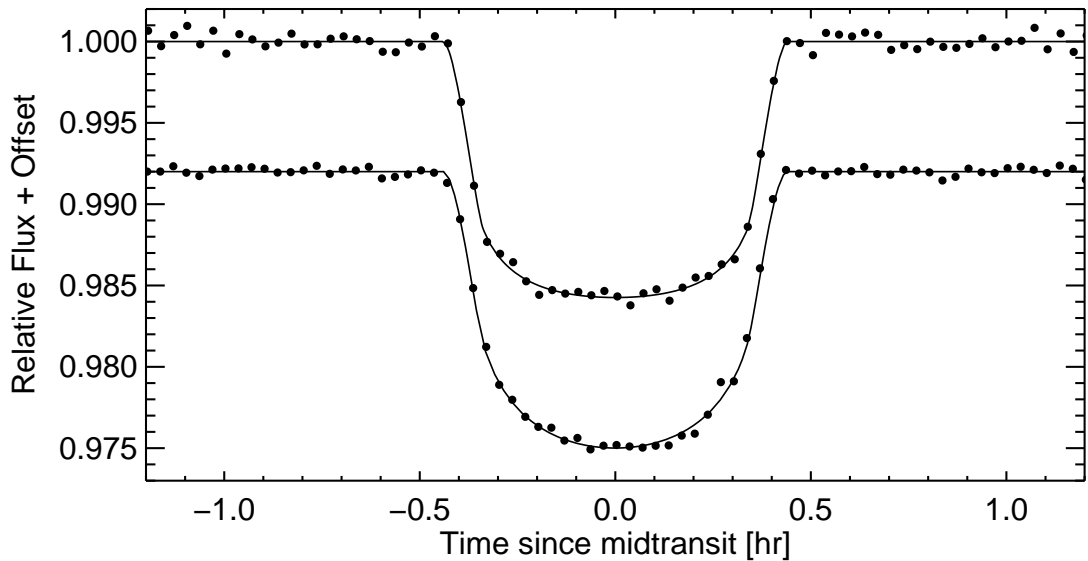}
           \caption{GJ~1214 composite transit light curves, in the $z'$-band (top) and $r'$-band (bottom), after merging all the data and resampling into two-minute bins. The black curves show the best-fitting model light curves.  The rms residual is 270 ppm for the $r'$-band light curve, and 560 ppm for the $z'$-band light curve.}
   \label{fig:comp}
\end{figure*}

Figures~\ref{fig:bigplots} and \ref{fig:closeup} show the minimum-$\chi^2$ solution (black curve) and the residuals for each night. Figure~\ref{fig:comp} shows the time-binned composite $z'$ and $r'$ light curves, with 2 minute sampling. The data were combined after correcting the depths for spot variability [such that $F \mapsto 1-(1-F)(1-\epsilon_j)$].  Table~\ref{tab:result} gives the results for each parameter, based on the $15.8\%$, $50\%$, and $84.2\%$ values of the cumulative posterior distribution for each parameter, after marginalizing over the other parameters. This table also gives some other parameters based on subsequent steps in the analysis, described in the sections to follow.

\begin{figure*}[h] %  figure placement: here, top, bottom, or page
   \centering
     \epsscale{1.15}
      \plotone{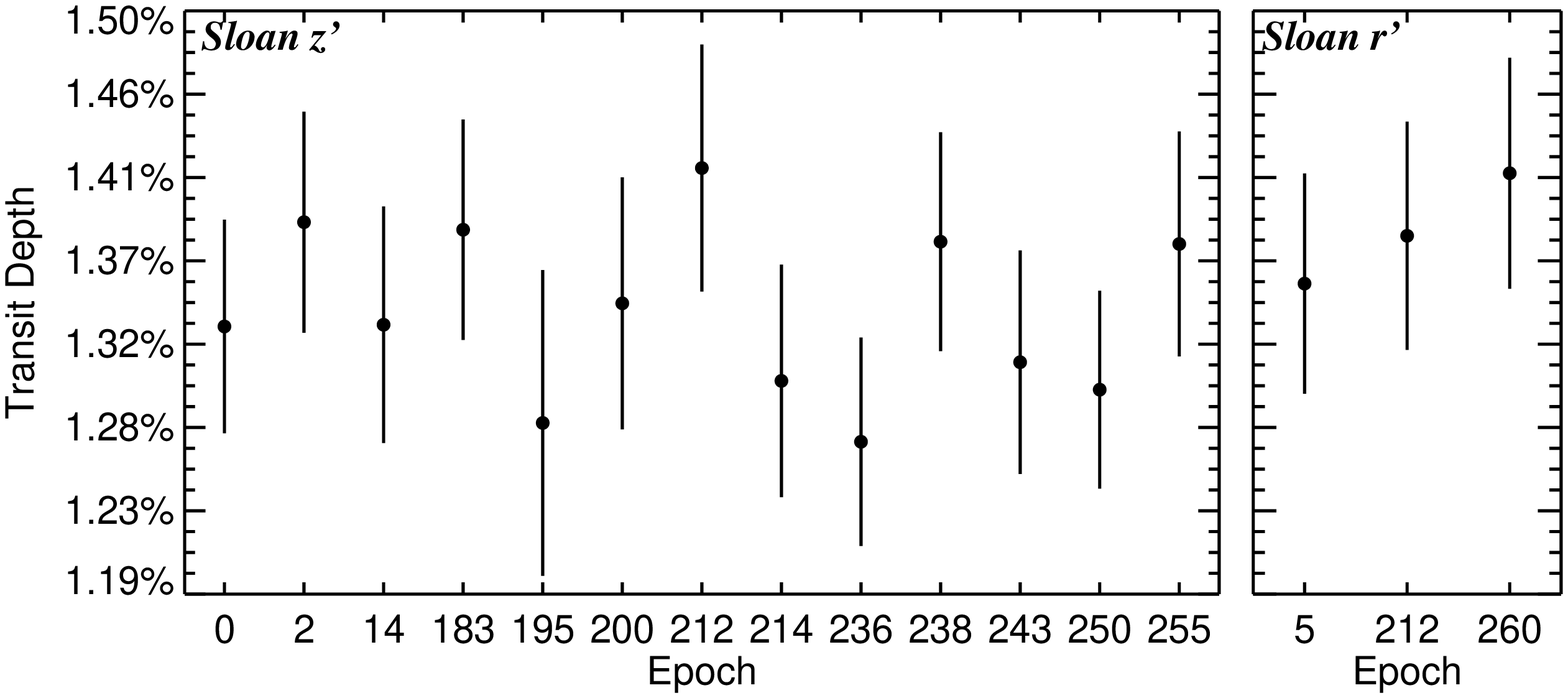}
      \caption{Transit light curve depth as a function of observation epoch {\new{(using values tabulated in Table~{\ref{tab:epoch}})}}.}
   \label{fig:depthvar}
\end{figure*}

\begin{figure*}[t] %  figure placement: here, top, bottom, or page
   \centering
     \epsscale{1.15}
      \plottwo{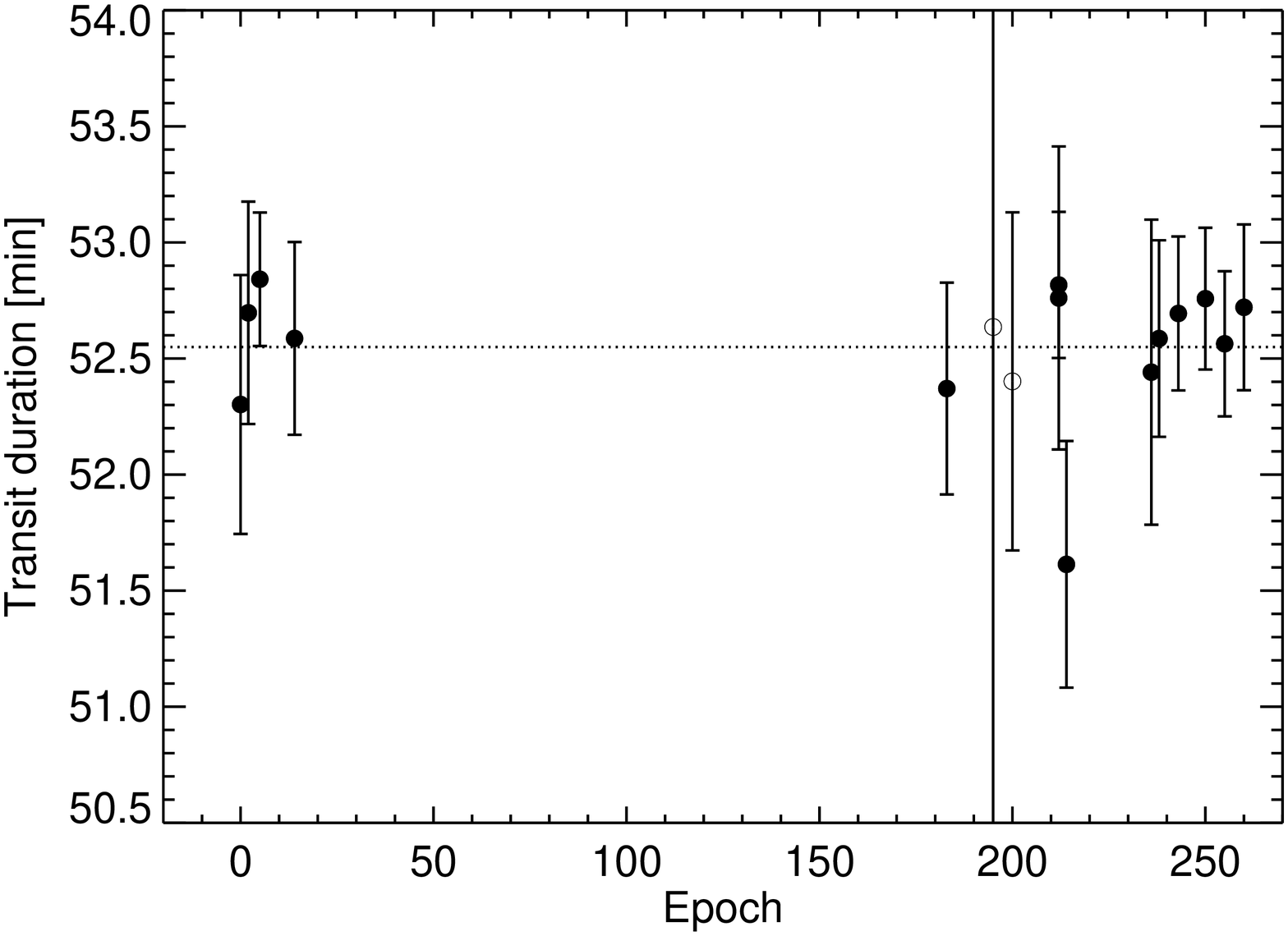}{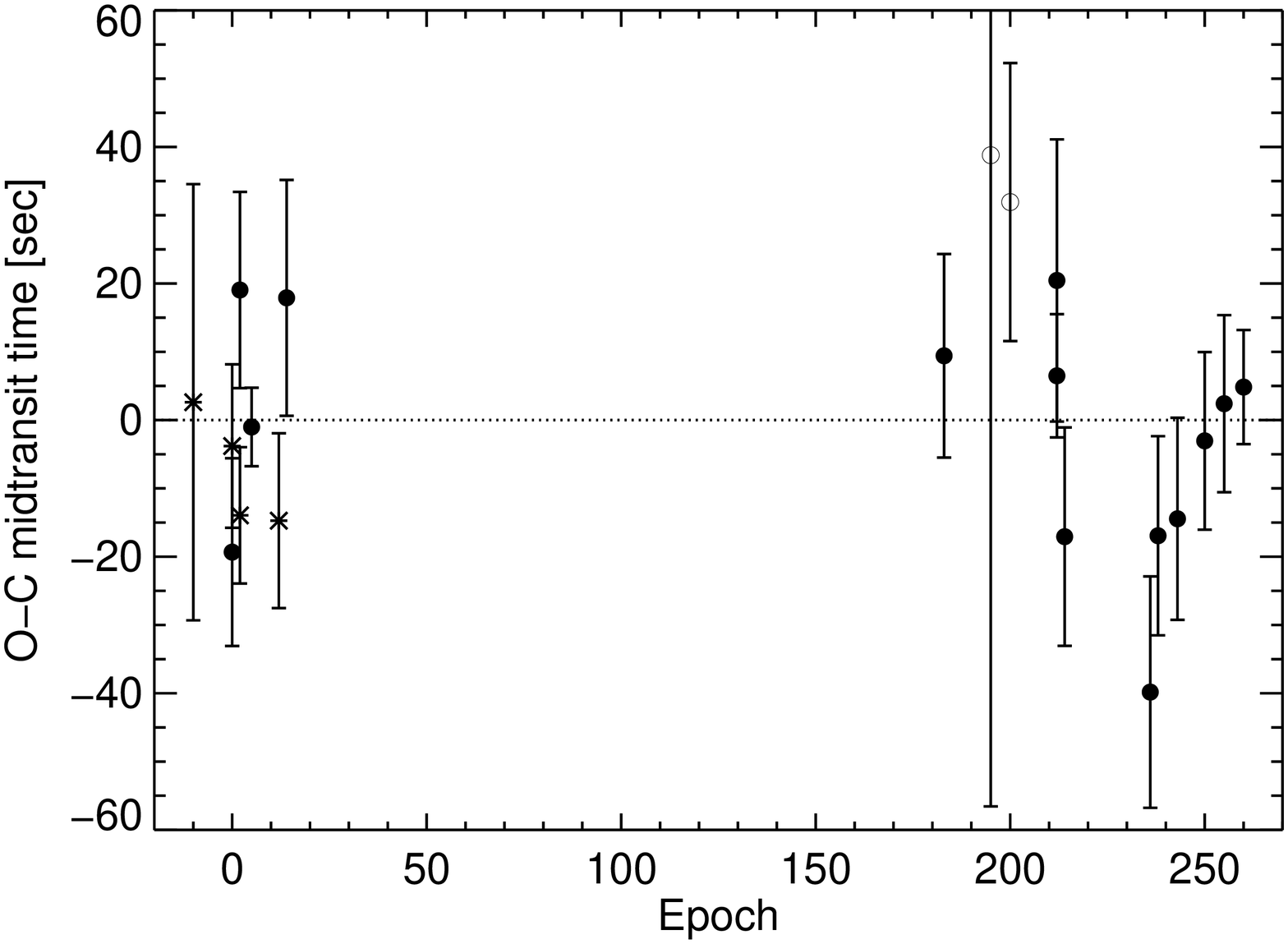}
      \caption{Transit durations and timing residuals. {\it Left.}---Total transit duration versus observation epoch.  {\it Right.}---Differences between the measured times of conjunction, and the best-fitting model assuming a constant period.  Only the complete transits presented in this work (solid circles) were used to derive the model parameters. Partial transits (open circles) were excluded from the fit.  The other plotted data (asterisks) are from Charbonneau et al.~(2009).}
  \label{fig:omc}
\end{figure*}

In order to assess any possible transit duration variations, we also performed a second analysis, in which the orbital inclination $i$ was allowed to be a free parameter specific to each night. Table~\ref{tab:epoch} gives the transit depth [$ \left(R_p/R_\star\right)^2/(1-\epsilon_j)$], time of conjunction, and total transit duration for each epoch. The depths are plotted as a function of epoch in Figure~\ref{fig:depthvar}. Figure~\ref{fig:omc} shows the durations, and the residuals from a linear ephemeris, as a function of epoch.

\begin{deluxetable*}{llclc}
\tablecolumns{5}
\tablewidth{0pt}
\tablecaption{System Parameters of GJ~1214 }
\tablehead{Parameter & \multicolumn{3}{l}{Value} & Comment} 
\startdata
\multicolumn{5}{c}{Transit Parameters} \\\hline  
Orbital period (day) & \multicolumn{3}{l}{$1.58040482\pm0.00000024$} & a  \\
Transit ephemeris (BJD$_{\rm TDB}$) & \multicolumn{3}{l}{2,454,980.7487955 $\pm 0.000045$} & a \\
Planet-to-star radius ratio, $R_p/R_\star$ & \multicolumn{3}{l}{$0.1161 \pm 0.0048$} & b \\
Orbital inclination, $i$ (deg)& \multicolumn{3}{l}{$90^{+0.0}_{-1.5}$} & c\\ 
Scaled semimajor axis, $a/R_\star$  & \multicolumn{3}{l}{$14.71^{+0.37}_{-0.33}$} & c\\
Total transit duration, $T_{\rm Total}$ (min) & \multicolumn{3}{l}{$52.73^{+0.49}_{-0.35}$} & \\
Transit ingress duration, $\tau$ (min) & \multicolumn{3}{l}{$5.80^{+0.41}_{-0.45}$} &  \\
Starspot-unaffected quantity, $(\rho_\star/\rho_\odot)^{-2/3}(R_p/R_\star)$ & \multicolumn{3}{l}{$0.017\pm0.001$} & d\\ 
$r'$ quadratic limb-darkening coefficients, ($u_r$,$v_r$) & ($0.49\pm0.07$, $0.48\pm0.22$)  & \\
$z'$ quadratic limb-darkening coefficients, ($u_z$,$v_z$) & ($0.17\pm0.12$, $0.64\pm0.30$) & \\\hline 
\multicolumn{5}{c}{Orbital Parameters} \\\hline
Radial velocity semiamplitude, $K_\star$ (m/s) & \multicolumn{3}{l}{$12.2 \pm 1.6$} & e \\
Orbital eccentricity, $e$ & \multicolumn{3}{l}{0} & fixed \\
{\new{Stellar projected rotational velocity, $v\sin i$ (km/s) }}& \multicolumn{3}{l}{ \new{$< 2.0$}} & {\new{e}} \\\hline
\multicolumn{5}{c}{Stellar Photometric Parameters \& Parallax} \\\hline
$J$ & \multicolumn{3}{l}{$9.750 \pm 0.024$} & e \\
$H$ & \multicolumn{3}{l}{$9.094 \pm 0.024$} & e \\
$K$ & \multicolumn{3}{l}{$8.872 \pm 0.020$} & e \\
Parallax, $\pi$ (mas) & \multicolumn{3}{l}{$77.2 \pm 5.4$} & e\\\hline
\multicolumn{5}{c}{Stellar \& Planetary Parameters} \\\hline
& Method A\dag & Comment & Method B\textdaggerdbl& Comment  \\ \cline{2-5}
Stellar mass, $M_\star$ ($M_\odot$) & $0.157\pm0.012$& & $0.156\pm0.006$ &   \\
Stellar radius, $R_\star$ ($R_\odot$) & $0.210 \pm 0.007$ &  &$0.179\pm0.006$ &   \\
Stellar mean density, $\rho_\star$ (g cm$^{-3}$) & $24.1\pm1.7$& f &  $38.4\pm2.1$ &   \\
Stellar effective temperature, $T_{\rm eff}$ (K) & -- & & $3170\pm23$&   \\
Stellar surface gravity, $\log g_\star$ (cgs) & $4.99 \pm 0.04$ & &$5.12\pm0.01$ &   \\
Stellar metallicity ([Fe/H]) & -- & & 0 & fixed  \\ \\
Planetary mass, $M_p$ ($M_\oplus$) & $6.45 \pm 0.91$& & $6.43\pm0.86$ &  \\
Planetary radius, $R_p$ ($R_\oplus$) & $2.65\pm0.09$& & $2.27 \pm 0.08$&  \\
Planetary mean density, $\rho_p$ (g cm$^{-3}$) & $1.89 \pm 0.33$ & & $3.03\pm0.50$&  \\
Planetary surface gravity, $g_p$ (m s$^{-2}$) & $9.0\pm 1.5$& g & $12.2\pm1.9$& 
\enddata
\label{tab:result}
\tablecomments{\dag Parameters calculated assuming an empirical luminosity-mass relationship (see \S~\ref{sec:radplanet:A} for details). \textdaggerdbl Parameters constrained to the Baraffe et al.\ (1998) stellar evolution isochrones (see \S~\ref{sec:iso} for details). a = Determined from a linear fit to the midtransit times listed in Table~\ref{tab:epoch}. b  = Determined by taking the square root of the variance weighted average of the depths listed in Table~\ref{tab:epoch} (see \S~\ref{sec:radrat} for details). c = Assuming epoch 236 coincides with the spot-free stellar surface. d = As calculated with Eqn.~\ref{eq:duration-radratio}. e = Charbonneau et al.~(2009). f = Estimated from transit parameters as $\rho_\star = (3 \pi)/(G P^2)(a/R_\star)^3$. g = Estimated from transit parameters as $(2\pi/P)K_\star (a/R_p)^2 (\sin i)^{-1}$ (Southworth et al.~2008). }
\end{deluxetable*}

\begin{deluxetable*}{llll}[h]
\tablecolumns{4}
\tablewidth{0pt}
\tablecaption{Epoch specific transit light curve results }
\tablehead{Epoch & Depth, $ (R_p/R_\star)^2 (1-\epsilon)^{-1}$ &Midtransit time (BJD$_{\rm TDB}$)& Duration (min)} 
\startdata
0 & $0.01332\pm0.00057$ & $2,454,980.74857\pm0.00015$& $52.30\pm0.56$\\
2 & $0.01388\pm0.00059$ & $2,454,983.90982\pm0.00016$ & $52.70\pm0.48$\\
5 & $0.01355\pm0.00059$ & $2,454,988.650808\pm0.000049$  &$52.84\pm0.29$ \\
14 &$ 0.01333\pm0.00063$&   $2,455,002.87467\pm0.00019$& $52.59\pm0.42$\\
183 & $0.01384\pm0.00059$& $2,455,269.96299\pm0.00016$&$52.37\pm0.46$ \\
195 & $0.01281\pm0.00081$& $2,455,288.9282\pm0.0011$&$52.64\pm6.23$\\
200 & $0.01345\pm0.00067$& $2,455,296.83013\pm0.00023$&$52.40\pm0.73$\\
212$^{\rm a}$ & $0.01416\pm0.00066$& $2,455,315.79485\pm0.00023$&$52.76\pm0.65$\\
212$^{\rm b}$ & $0.01380\pm0.00061$& $2,455,315.794693\pm 0.000080$ & $52.82\pm0.31$\\
214 & $0.01303\pm0.00062$ &$2,455,318.95523\pm0.00017$ &$51.61\pm0.53$ \\
236 & $0.01271\pm0.00055$&$2,455,353.72387\pm0.00018$ &$52.44\pm0.66$ \\
238 & $0.01377\pm0.00058$&$2,455,356.88495\pm0.00015$ & $52.59\pm0.42$\\
243 & $0.01313\pm0.00059$& $2,455,364.78700\pm 0.00015$& $52.69\pm0.33$\\
250 & $0.01299\pm0.00053$&$2,455,375.84997\pm0.00013$ & $52.76\pm0.31$\\
255 & $0.01376\pm0.00060$& $2,455,383.75205\pm0.00013$&$52.56\pm0.31$\\
260 & $0.01414\pm0.00061$& $2,455,391.654105\pm0.000059$ & $52.72\pm0.36$\\ \hline \\
Mean$^{c}$ & $0.01348\pm0.00011$&  &
\enddata
\label{tab:epoch}
\tablenotetext{a}{Sloan $z'$}
\tablenotetext{b}{Sloan $r'$}
\tablenotetext{c}{Variance weighted mean.}
%\tablerefs{}
\end{deluxetable*}

\section{The planet-to-star radius ratio} \label{sec:radrat}

\begin{figure*} %  figure placement: here, top, bottom, or page
   \centering
     \epsscale{1.15}
      \plotone{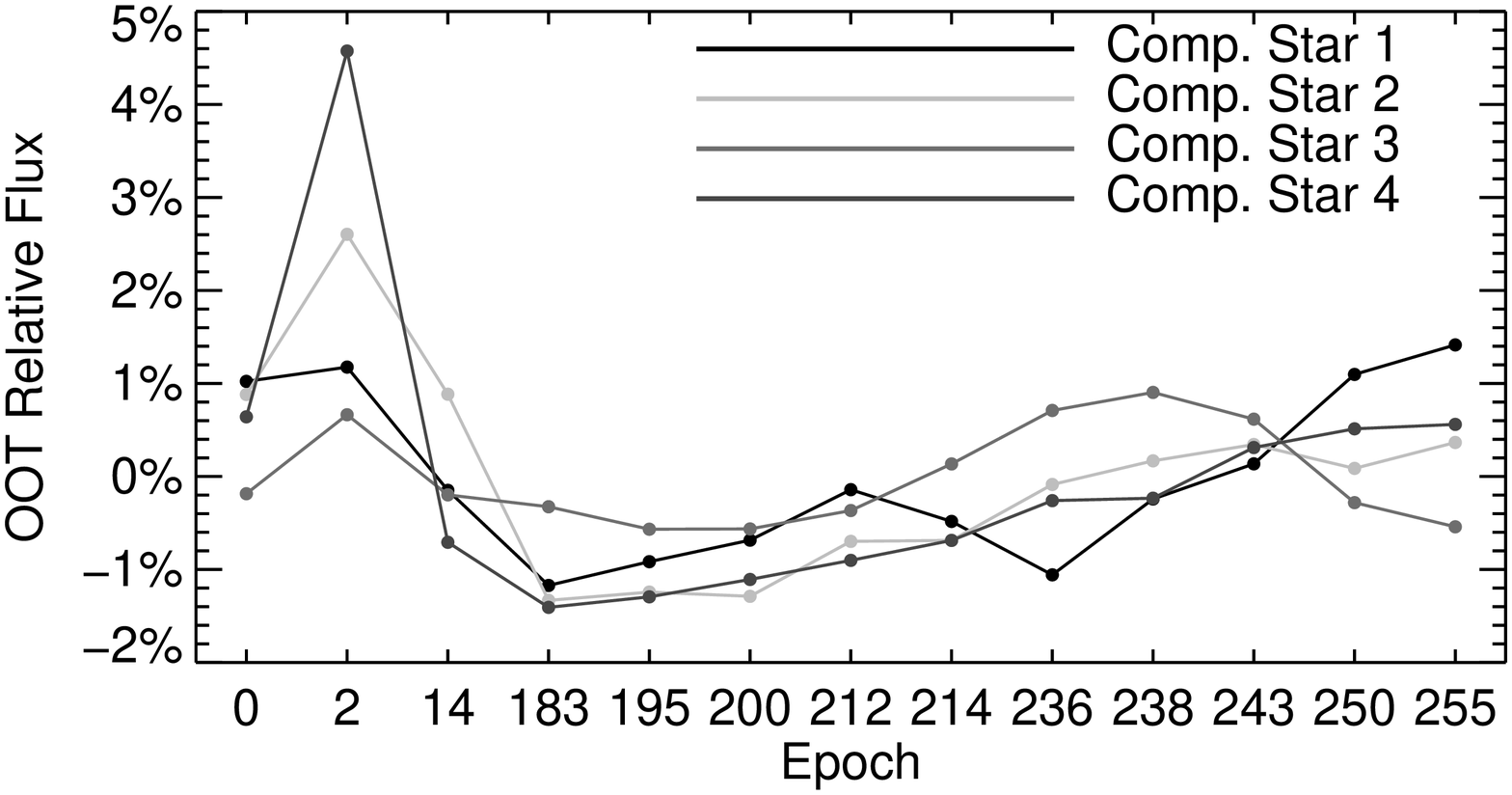}
      \caption{Out-of-transit variability of GJ~1214.  Plotted is the flux (measured out-of-transit) of GJ~1214, relative to four
different comparison stars. }
   \label{fig:compstars}
\end{figure*}

Despite our concerns about the effects of starspots on the light curves, we did not find any significant variation of the transit depth or and duration with time, and we did not find any significant departure from a linear ephemeris. In addition, the out-of-transit flux of GJ~1214 measured directly from the images was constant to within a few percent over the course of our observations (see Figure~\ref{fig:compstars}). There was apparently a decline by 1--2\% between 2009 and 2010, followed by a gradual 1--2\% rise throughout our 2010 observations. This should have been accompanied by 1--2\% variations in the measured transit depth, which is beneath our typical measurement precision of 5\%, and hence it is not surprising that no such trend was observed.

Based upon this finding, one way to estimate the planet-to-star radius ratio is to assume that spots have not significantly affected the transit depth. We may then calculate the radius ratio based on the variance-weighted average of the 16 transit depths $\delta_j \equiv (R_p/R_\star)_j^2 (1-\epsilon_j)^{-1}$:
\begin{eqnarray}
	\frac{R_p}{R_\star} &  = &\sqrt{ \frac{1}{\sum \sigma_{\delta_j}^{-2}} \sum_{j=1}^{16} \frac{\delta_j}{ \sigma_{\delta_j}^2}} \\ \nonumber
						& = & 0.11610\pm0.00048.
\end{eqnarray}
where the average is taken over all epochs $j$.  This result for the radius ratio is slightly smaller than that reported in the discovery paper ($0.11620 \pm 0.00067$; Charbonneau et al. 2009), but the two results are consistent to within the uncertainties.

Due to time variability of starspots at the 1--2\% level, this estimate is subject to a bias of a few percent, in the sense that the planet-to-star radius ratio may actually be a few percent {\it smaller} than $0.11610$.  An even more conservative stance would recognize that we cannot exclude even larger effects due to the time-independent component of the starspot coverage (if, for example, the poles of the star were persistently darker than the rest of the photosphere).  In that sense we can only place an upper bound on the radius ratio: $R_p/R_\star \le 0.1161$ at 95\% confidence {\new{(based on the marginalized posterior distribution produced by the MCMC algorithm).}}

It should also be noted that the $r'$-band transit depth ($1.383\% \pm 0.035\%$) was found to be slightly larger than the $z'$-band transit depth ($1.340\% \pm 0.017\%$), although the difference is only at the 1--2$\sigma$ level. A deeper $r'$-band transit is expected if cool starspots are affecting the results. In contrast, models of the atmosphere of GJ~1214b generally predict that the transit depth should have the {\em opposite} wavelength dependence, with a deeper $z'$-band transit {\new{ (see, e.g., Fig.~3 of Miller-Ricci \& Fortney 2010)}}. Those who would attempt to attribute any slight wavelength-dependence of the transit depth to the planetary atmosphere should beware of the confounding effects of starspots.

\section{The radius of the planet}  \label{sec:radplanet}

In order to understand the structure and atmosphere of GJ~1214b, we want to know its radius, rather than just the planet-to-star radius ratio. This requires some externally derived estimate of the stellar radius, or mass, or both. \ron{As we will describe, w}e have investigated two different pathways to the planetary radius, relying on different assumptions, and found them to give inconsistent results. The results of both of these methods, including a number of derived stellar and planetary parameters, are given in Table~2. This inconsistency had already been noted by Charbonneau et al.~(2009), but here we delve further into the details and discuss possible resolutions.

\subsection{Method A: Empirical mass-luminosity relation,
and transit-derived stellar mean density } \label{sec:radplanet:A}

In the first method, the stellar mass is estimated based on its observed luminosity (parallax, and apparent magnitude). Then, the stellar radius is found by combining the stellar mass with the mean stellar density $\rho_\star$ derived from the transit light curve,
\begin{eqnarray}
\rho_\star = \frac{3 \pi}{G P^2} \left(\frac{a}{R_\star}\right)^3 \left(1+\frac{M_p}{M_\star}\right)^{-1} \label{eq:dens}
\end{eqnarray}
which, for $M_p \ll M_\star$, is only a function of the photometrically-determined parameters $a/R_\star$ and $P$.

We begin with the $K$-band mass-luminosity function of Delfosse et al.\ (2000). For GJ~1214, with a parallax $\pi = 77.2\pm 5.4$~mas and apparent $K$ magnitude $8.782\pm 0.020$, we used the polynomial function given by Delfosse et al.\ (2000) to estimate $M_\star = 0.157\pm0.012$ $M_\odot$, where the uncertainty is based only on the propagation of errors in $\pi$ and $K$. There should be additional uncertainty due to the intrinsic scatter in the mass-luminosity relation, but this intrinsic uncertainty was not quantified. Charbonneau used the same method and reported $M_\star = 0.157\pm0.019$ $M_\odot$.

Next, using $a/R_\star = 14.71\pm0.35$, as derived from our ensemble of light curves, we found the mean stellar density to be $\rho_\star = 24.1\pm1.7$ g/cm$^3$. By combining $M_\star$ and $\rho_\star$ we found the stellar radius to be
\begin{equation}
R_\star = 0.210 \pm 0.007 R_\odot~~{\rm (method~A)}.
\end{equation}

Finally, assuming that our transit depths were not affected by starspots (i.e., $\epsilon_j \equiv 0$) we used the radius ratio derived from our observations, $R_p/R_\star = 0.11610\pm0.00048$, to compute the planetary radius:
\begin{equation}
R_p = 2.650\pm0.090~R_\oplus~~{\rm (method~A)}.
\end{equation}
This is in agreement with that found by Charbonneau et al.\ (2009), who found $R_p=2.678\pm 0.130~R_\oplus$.  Taking the more conservative stance that spots may have biased the radius ratio measurement, we may set an upper bound on the radius of the planet: $R_p \le 2.81$~$R_\oplus$ with 95\% confidence.
 
\subsection{Method B: Stellar evolutionary models} \label{sec:iso}

The mass-luminosity relationship of Delfosse et al.~(2000) has the virtue of being highly empirical, as it is based on observations of astrometrically resolved, detached eclipsing binaries with measured parallaxes. However, with only 16 systems as inputs, the empirical relation cannot be expected to account for all the relevant variables such as age and metallicity. An alternate approach is to trust theoretical models of stellar evolution, which predict the radius of a star as a function of its mass, age, and metallicity, and are calibrated by observations when possible.

To try this approach, we used the stellar evolutionary models of Baraffe et al.\ (1998), which are presented as a series of stellar parameters as a function of age (isochrones) for a given metallicity. Since the metallicity of GJ~1214 is unknown we assumed a solar metallicity. We assigned a likelihood ${\cal L}(j) \propto \exp[-\chi^2(j)/2]$ to every isochrone entry $j$ such that
\begin{equation}
	\chi^2(j) = \sum_k \frac{\Delta p_k^2}{\sigma_{p_k}^2}
\end{equation}
where $\{p_k\}$ is the set of stellar parameters that is subject to observational constraints (e.g.\ absolute magnitudes in
certain bandpasses), $\sigma_{p_k}$ are the corresponding 1$\sigma$ uncertainties, and $\Delta p_k$ are the differences between the isochrone values and the ``observed'' values.  Then, the most likely value for a given stellar parameter $p$ (such as stellar radius) was found by taking a weighted average over all the isochrone points,
\begin{eqnarray}
	\langle p \rangle & =& \frac{\sum_j p(j) {\cal L}(j)}{\sum_j {\cal L}(j)}.
\end{eqnarray}
The uncertainty in this parameter is taken to be the square root of the weighted variance, i.e., $\sigma_p = \sqrt{\langle p^2 \rangle - \langle p \rangle^2}$ where averages are taken as defined above.\footnote{This approach is closely related to the technique advocated by Torres et al.\ (2008).  The relatively coarse sampling of the Baraffe et al.\ (1998) isochrones precluded the use of more sophisticated numerical integration techniques such as those described by Carter et al.\ (2009).}  We used this technique with constraints based on the observed apparent magnitudes in the $J$, $H$, and $K$ bandpasses and the parallax, all from Charbonneau et al.~(2009).  The results were $M_\star = 0.131 \pm 0.044$ $M_\odot$ and $R_\star = 0.191 \pm 0.019$ $R_\odot$.

This method gives a significantly lower mass than the first method. This difference is almost entirely due to the additional degree of freedom of the stellar age: the theoretical evolutionary tracks allow for GJ~1214 to have any age, while the Delfosse et al.~(2000) relation is a consensus result based on a fit to stars of varying age, most of which are probably older than a few Gyr.

This naturally raises the question: what is known about the age of GJ~1214?  As noted by Charbonneau et al.~(2009), and confirmed in this work, the optical light variations of GJ~1214 are only a few percent in amplitude, suggestive of a more mature star. Moreover, age estimates based on the star's space velocity and the estimated stellar rotation period also favor an older star, leading Charbonneau et al.~(2009) to estimate an age of 3--10~Gyr.  To account for this evidence, we repeated our isochrone analysis but with a prior constraint on the stellar age.  Specifically, we used a Gaussian prior on $\log_{10}$(age) with a median value of 10 and a standard deviation of 0.4.  With this extra constraint, the stellar mass was found to be $M_\star = 0.156\pm0.006$ $M_\odot$, in excellent agreement with the results of the first method.

This substantiates our claim that the main difference between the two methods of estimating the stellar mass is the treatment of stellar age.  However, there remains a significant discrepancy in the stellar radius. The evolutionary models with the ``old age'' prior give
\begin{equation}
R_\star = 0.179 \pm 0.006 R_\odot~~{\rm (method~B)}.
\end{equation}
which is about 15\% smaller than the result of the first method.
Again taking the radius
ratio to be $R_p/R_\star = 0.11610\pm0.00048$, the planetary radius is
\begin{equation}
R_p = 2.270\pm0.080~R_\oplus~~{\rm (method~B)}.
\end{equation}

\subsection{Possible resolutions of the two methods}

To summarize the preceding results, we have investigated two methods of estimating the stellar mass and radius based on the available information. The two methods can be made to agree on the stellar mass, but they refuse to agree on the stellar radius. The first method is based on an empirical mass-luminosity relation, and the stellar mean density derived from the transit light curve. The second method is based on stellar evolutionary models constrained by the infrared absolute magnitudes and an assumed age $\gtrsim$1~Gyr. The first method gives a stellar radius that is larger by 15\%, which is approximately 4.5 times larger than the internal uncertainty in either estimate. The discrepancy is even starker if one compares the mean stellar density derived from the transit light curve ($24.1\pm1.7$~g~cm$^{-3}$) with that derived from the stellar evolutionary models ($38.4\pm2.1$~g~cm$^{-3}$), as they disagree by 7$\sigma$.

It has long been appreciated that stellar evolutionary models have trouble predicting the masses and radii of M dwarfs in detached eclipsing binary systems (see, e.g., Hoxie 1973, Lacy 1977), in the sense that the models tend to underpredict the observed radius for a given mass by about 10\% (Ribas 2006). It was for this reason that Charbonneau et al.~(2009) discounted the results based on the evolutionary models, and why subsequent authors have done the same in their discussions of GJ~1214.  However, it has recently been argued that the failings of the evolutionary models are confined to the mass range 0.3--0.7~$M_\odot$, and even more specifically to stars that have been tidally spun up (and made more active) by a close stellar companion (see, e.g., Torres et al.~2006, L\'opez-Morales 2007, Morales et al.~2008). Stars below a threshold mass of $\approx$0.32~$M_\odot$ are expected to be fully convective, and seem to be well described by the evolutionary models (see, e.g., L\'opez-Morales 2007, Demory et al.~2009, Vida et al.\ 2009).  To the extent this is true, we would not expect GJ~1214---a single, low-activity star of mass $<$0.3~$M_\odot$---to be affected by the problems that plague the theories of earlier-type M dwarfs in close binary systems.

We are thereby motivated to seek alternative resolutions to the discrepancy between the two methods of estimating the mass and radius of GJ~1214.  Of course there is always the possibility that a key input such as the parallax or infrared magnitude is faulty, but in the sections to follow we discuss some possible resolutions in which all of the data are taken at face value.

\subsubsection{A metal-poor star?}

We have worked exclusively with solar metallicity isochrones. Lower-metallicity isochrones generally predict a larger radius for a given mass, in the relevant region of parameter space.  Schlaufman~(2010) have presented a simple method to estimate the metallicity of an M dwarf based on its observed $V$ and $K$ magnitudes. For GJ~1214, his method gives [Fe/H]$=-0.16$, suggesting that the star is only moderately metal-poor. This value for the metallicity would not affect the theoretically predicted radius by enough to resolve the discrepancy.  L\'opez-Morales (2007) found that the Baraffe et al.~(1998) isochrones predict a variation in stellar radius of only about 3\% for metallicities ranging from 0.0 to $-0.5$.  Moreover, for a sample of low-mass stars with $-0.5 < $[Fe/H]$ < 0$, Demory et al.\ (2009) showed that there is no significant correlation between measured metallicity and the fractional difference between the measured stellar radius and that found assuming solar metallicity. Therefore, while it is possible that GJ~1214 is a metal-poor star, we do not consider this to offer a likely resolution of the discrepancy we have noted between the two methods of estimating the stellar radius.

\subsubsection{A young star?}

If the star were relatively young and still contracting onto the main sequence, then the evolutionary models would predict a larger radius, relieving the discrepancy.  To investigate this possibility we repeated our isochrone analysis, but this time with a flat prior on the stellar age and a Gaussian prior on the stellar mean density to conform with the transit light curve analysis.  The result was that the age of the star must be about 100~Myr. This would conflict with the evidence for an older age, namely, the observed lack of strong chromospheric activity and the kinematics (see \S~\ref{sec:iso}). There is also the low probability that we would happen to observe this star at such an early phase of its long life. For these reasons, a young age for GJ~1214 does not seem to be a promising solution.

\subsubsection{A starspot-corrupted estimate of $a/R_\star$?}

It is conceivable that unidentified starspot anomalies have significantly biased our estimate of $a/R_\star$ from the transit light curves (see \S~\ref{sec:model:spots}). Could this be responsible for the 7$\sigma$ discrepancy in $\rho_\star$ between the two methods of characterizing the star? If this were the case, then the more trustworthy estimate of $\rho_\star$ would be the value from the isochrone analysis.

This seems unlikely, partly because the residuals to our transit light curves do not display any anomalies beyond the two that we have already identified, and partly because in this scenario it would be difficult to understand the collection of transit depths.  Specifically, we can use the value of $\rho_\star$ from the evolutionary models as an input to Equation~(\ref{eq:duration-radratio}), to derive the planet-to-star radius ratio. The result is $R_p/R_\star = 0.15\pm0.01$.  This conflicts with the upper limit $R_p/R_\star \le 0.1161$ that we derived in \S~\ref{sec:radrat}, under the assumption that the transit depths are affected by cool spots on the stellar disk.  Thus, one would have to suppose that nearly all of the transit depths were biased to {\it smaller} values by numerous starspot crossings throughout the transits.  We do not consider a conspiracy of starpot anomalies to be a satisfactory solution to the problem.

\subsubsection{Is the orbit eccentric?}

If the orbit of GJ~1214b is not circular, then the parameter that is being determined by the light curve analysis is not $a/R_\star$ and our application of Equation~(\ref{eq:dens}) is erroneous. The correct procedure must take into account the speed of the planet at inferior conjunction, which is a function of the eccentricity and argument of pericenter (see, e.g., Eqns.~16 and 27 of Winn 2010 or Kipping 2010).
The end result is that the mean density of the planet would be modified as follows
\begin{equation}
\rho_\star = \rho_{\star,~{\rm circ}} \left( \frac{\sqrt{1-e^2}}{1+e\sin\omega} \right)^3,
\end{equation}
where $\rho_{\star,~{\rm circ}}$ is the mean density that is calculated under the assumption of a circular orbit.  Therefore, it might be possible to reconcile the two different methods for estimating the stellar density, for suitable choices of $e$ and $\omega$.

The orbital eccentricity has been assumed to be zero, because of the expectation that tidal dissipation has damped out any initial eccentricity to $10^{-3}$ or below.  However, the published RV data only permit a coarse upper limit of {\new{$e$}}$<$0.27 with 95\% confidence (Charbonneau et al.~2009). To investigate the possibility of an eccentric orbit, we assumed that the isochrone-based estimate of the mean stellar density ($38.4\pm2.1$ g~cm$^{-3}$) is accurate.  We then derived the constraints on the orbital eccentricity that would
be required for consistency with the transit light curve analysis.
The minimum eccentricity is obtained for the case $\omega=90\arcdeg$,
corresponding to transits occurring at pericenter.  In that case, $e =
0.138 \pm 0.034$.  Solutions with $e<0.27$ can be found for values of
$\omega$ between $25\arcdeg$ and $155\arcdeg$.

The only objection is that tidal dissipation should have damped out this eccentricity. The characteristic circularization timescale is $\tau_c\sim 10$~Myr assuming a tidal quality factor $Q_p=100$ for the planet, a rough order-of-magnitude estimate for a solid planet. This is much shorter than the estimated stellar age. For an icy body like Neptune with $Q_p \sim 10^4-10^5$, the circularization timescale would be 1--10~Gyr, which is at least not overwhelmingly shorter than the system age. The mechanisms and timescales for tidal dissipation are not understood from first principles, and are poorly constrained by observations. And it should be kept in mind that all of the known transiting Neptune-like planets (GJ~436, Kepler-4, and HAT-P-11) all have significantly eccentric orbits.  For these reasons, of all the possibilities we have discussed, we find an eccentric orbit to be the most attractive solution to the radius
discrepancy problem.

\section{Analysis of Midtransit times: \\Limits on perturber mass} \label{sec:timing}

We recalculated the transit ephemeris utilizing 14 of the 16 midtransit times given in Table~\ref{tab:epoch}.  We excluded the two midtransit times inferred from partial transit light curve data (epochs 195 and 200), out of concern that the second-order airmass correction could not be determined as well in those cases. We fitted a linear function of transit epoch $E$,
\begin{eqnarray}
	T_c(E) & = & T_c(0)+E P.
\end{eqnarray}
The results were $T_c(0) = 2454980.7487955 \pm 0.000045$ (BJD$_{\rm TDB}$) and $P = 1.58040482 \pm 0.00000024$ days and the fit had $\chi^2 = 19.2$ with 12 degrees of freedom. Formally the fit is inconsistent with the linear model with 90\% confidence, but we do not consider this to be a significant detection of timing anomalies.  The right panel of Fig.~\ref{fig:omc} shows the residuals to the linear fit. This plot also shows four data points from Charbonneau et al.\ (2009), which are consistent with the linear model.

Since \ron{there} is no clear evidence for transit timing variations (TTVs), we may use the timing data to place upper limits on the mass of a hypothetical second planet that would perturb the orbit of the transiting planet. To do so, we used an implementation of the algorithm advocated by Steffen \& Agol (2005).  We explored the parameter space of the hypothetical
perturber's mass and orbital period and phase. We assumed the orbits 
of the perturber and GJ1214 were circular and coplanar. 
Regions in parameter space yielding dynamical 
instability, determined following the prescription by Barnes 
\& Greenberg (2006), were not sampled. For each orbital period 
ranging from 0.3--8 days and sampling all orbital phases, the 
mass of the perturber was increased until the computed transit 
times would fit our data $\Delta \chi^2 = 9$ worse than a linear 
ephemeris; this mass is interpreted as a 99.7\% confidence 
upper-limit.

\begin{figure*}[t] %  figure placement: here, top, bottom, or page
   \centering
     \epsscale{0.95}
      \plotone{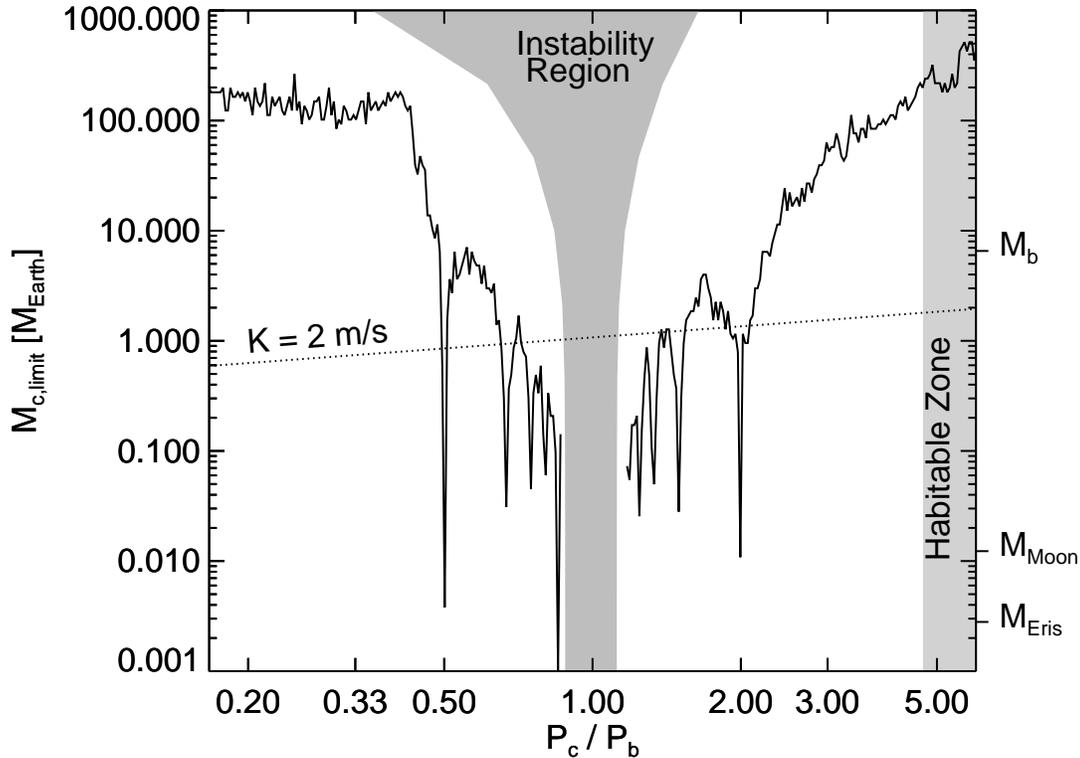}
      \caption{Upper mass limits for a hypothetical perturber as a function of perturber period {\new{normalized to the period of GJ~1214b}}, $P_c${\new{$/P_b$}}.   }
   \label{fig:masslimits}
\end{figure*}

Fig.~\ref{fig:masslimits} shows the constraints on the perturber mass as a function of period ratio, as determined from this analysis. For reference, on the right-hand axis we have included the masses of GJ~1214b (6.6~$M_\oplus$), the Moon (0.01~$M_\earth$), and Eris (0.003~$M_\earth$). In this plot we have also indicated the zones of dynamical instability, and of potential habitability (equilibrium temperature between 273--373~K, for an assumed albedo of zero). We have also plotted a line corresponding to an RV amplitude of 2~m~s$^{-1}$, probably the best achievable detection limit based only on RV observations for the near future.

The mass constraints on the perturber are more restrictive near the mean-motion resonances and most restrictive at the low-order resonances, particularly for the interior and exterior 2:1 resonances.  For example, a perturber at the interior 2:1 resonance having mass near that of Eris would have induced detectable TTVs with the present data.

\section{Discussion} \label{sec:disc}

One of our goals in this study was to improve on the estimates of the basic parameters of GJ~1214b. Yet despite having undertaken many high-precision observations of transits, we have not significantly improved on the estimate of the planetary radius. In fact\ron{,} our analysis has led us to conclude that the radius is even more uncertain than had been recognized previously. This is for two reasons. First, the clear evidence for starspots in our most precise light curves has caused us to consider the possible effects of stellar activity on the analysis of transit photometry. Second, and more importantly, we have found a significant disagreement between two methods of estimating the stellar and planetary dimensions, with no good reason why either one should be disregarded or considered less trustworthy. Both of these complications were known prior to this work {\new{(e.g., Charbonneau et al. 2009)}}, but we have brought them into {\new{sharper}} focus.

The problem of star spots can be mitigated by observing in longer-wavelength bandpasses.  This is because the flux contrast between two blackbodies of different temperatures is a decreasing function of wavelength.  {\new{For example, if we assume}} the particular starspots that produced anomalies in our data are approximately the same size as the planet, then they are $\approx$150~K cooler than the stellar photosphere (see \S~\ref{sec:spotcross} for details).  This corresponds to a surface-brightness ratio of about 0.67 between the spots and the surrounding \ron{photosphere}, at an observing wavelength of 0.6~$\mu$m.  At 3.5~$\mu$m, the surface-brightness ratio would be about 0.91, representing a smaller contrast and correspondingly smaller starspot-induced effects.

The problem of the stellar radius will be more difficult to solve, and there is much at stake. The mean planetary density could be $1.89 \pm 0.33$~g~cm$^{-3}$ or $3.03 \pm 0.50$~g~cm$^{-3}$, depending on which route is taken to estimate the stellar radius. The lower density would imply that the planet must have a dense gaseous atmosphere, for which there are many intriguing possible origins and compositions (Rogers \& Seager 2010, Miller-Ricci \& Fortney 2010). In contrast, the higher density could be consistent with a solid planet with a very thin (or nonexistent) atmosphere. We have discussed several possible resolutions of this discrepancy, and argued that the most attractive possibility is that the planet has an eccentric orbit, $e \approx 0.14$.  This hypothesis can be tested by gathering additional RV data, or by measuring the time of occultation of the planet by the star.  Neither task will be easy.  Assuming the minimum detectable eccentricity to vary inversely as the square root of the number of RV data points, one would need approximately 4 times as many data points (with the same precision as the existing data) for a 2$\sigma$ detection of an eccentricity of 0.14.  Likewise, the occultation depth is expected to be smaller than 0.1\% at 3.5~$\mu$m, the region accessible to the {\it Spitzer} space telescope, which probably offers the best prospects for such a detection.

This experience with GJ 1214 invites some broader remarks about the suitability of M dwarf stars as targets in surveys for transiting planets. The {\new{advantages of M dwarfs have been discussed}} by Nutzman \& Charbonneau (2009), among others. Their smaller radii and masses allow for smaller planets to be detected, at a given signal-to-noise ratio. They are numerous in our Galactic neighborhood. Transits of planets in the habitable zone in particular are more probable {\new{and, hence}}, more frequent than they are around higher-mass stars. These are decisive advantages, fully justifying the ongoing efforts that concentrate exclusively on M dwarfs. However, there are two disadvantages of low-mass stars. They tend to have larger spots and an overall higher level of stellar activity than more Sun-like stars, which will interfere with the measurement of the basic transit parameters (c.f.\ Seager \& Deming 2009). And, it will be harder to obtain reliable estimates of the stellar mass and radius, because of possible limitations of stellar evolutionary models in the range 0.3--0.7~$M_\odot$ and, more generally, because of our more limited empirical knowledge of the lowest-mass stars.

\end{document}